\documentclass[journal, onecolumn, draftcls,12pt]{IEEEtran}
\usepackage{amssymb,amsmath,epsfig,graphicx,theorem}
\usepackage{amsfonts}
\usepackage{bm}
\newtheorem{Theo}{Theorem}
\newtheorem{Lem}{Lemma}
\newtheorem{Cor}{Corollary}

\usepackage{epsfig,rotating,setspace,latexsym,amsmath,epsf,amssymb,bm}
\usepackage{cite}

\IEEEoverridecommandlockouts

\title{The Gaussian Multiple Access Diamond Channel}

\author{\IEEEauthorblockN{Wei Kang \qquad \qquad Nan Liu \quad \quad Weiwei Chong\\}
\IEEEauthorblockA{School of Information Science and Engineering\\
Southeast University, Nanjing, Jiangsu, P. R. China, 210096\\
\emph{wkang@seu.edu.cn \qquad \qquad nanliu@seu.edu.cn} \quad \quad
\emph{chongww@yeah.net}}
\thanks{This paper was presented in part at IEEE International Symposium on Information Theory (ISIT), 2011.  This work is partially
supported by the National Basic Research Program of China
(973 Program 2012CB316004), the National Natural Science Foundation of China
under Grants $61271208$, $61201170$  and $61221002$,
the Research Fund of National Mobile Communications Research Laboratory,
Southeast University (No. 2014A02),
the Project-sponsored by SRF for ROCS, SEM and Qing Lan Project.}}

\begin{document}
\maketitle

\begin{abstract}
In this paper, we study the capacity of the diamond channel. We
focus on the special case where the channel between the source node
and the two relay nodes are two separate links with finite
capacities and the link from the two relay nodes to the destination
node is a Gaussian multiple access channel. We call this model the
Gaussian multiple access diamond channel. We first propose an upper
bound on the capacity. This upper bound is a single-letterization of
an $n$-letter upper bound proposed by Traskov and Kramer, and is
tighter than the cut-set bound. As for the lower bound, we
propose an achievability scheme
based on sending correlated codes through the multiple access
channel with superposition structure. We then specialize this achievable rate to
the Gaussian multiple access diamond channel. Noting the similarity
between the upper and lower bounds, we provide sufficient and necessary conditions that a
Gaussian multiple access diamond channel has to satisfy such that
the proposed upper and lower bounds meet. Thus, for a Gaussian
multiple access diamond channel that satisfies these conditions, we
have found its capacity.
\end{abstract}

\emph{keywords:} Correlated codes, diamond channel, Gaussian
channel, multiple access channel

\newpage

\section{Introduction}
The diamond channel was first introduced by Schein in 2001
\cite{Schein:2001}. It models the communication from a source node
to a destination node with the help of two relay nodes. The channels
between the source node and the two relay nodes form a broadcast
channel as the first stage and the channels between the two relay
nodes and the destination node form a multiple access channel as the
second stage. The capacity of the diamond channel in its general
form remains unknown. Achievability results were proposed in
\cite{Schein:2001}, while for the general diamond channel, the best
known converse result is still the cut-set bound \cite{Cover:1991}.
Capacity has been found for some special classes of discrete diamond channels
in \cite{Kang:2011b, Tandon:2010}.
For the Gaussian diamond channel, the capacity is approximated within $1$ bit \cite{Avestimehr:2011}. For the Gaussian $N$-relay diamond channel, a uniform approximation of the capacity has been obtained in \cite{Niesen:2013}, where bursty amplify-and-forward was proposed as the achievability scheme and simultaneous optimization over all possible cuts was used for the converse.

The problem of sending correlated codes through a multiple access
 channel was studied in \cite{Ahlswede:1983}. This channel
model can be regarded as a special class of the diamond channel
where the broadcast channel between the source node and the two
relay nodes are two separate links of finite capacities.
We call this channel model \emph{the multiple access diamond
channel}. Achievability results for the discrete multiple access
diamond channel were proposed in \cite{Ahlswede:1983,Traskov:2007}.
In \cite{Traskov:2007}, an uncomputable $n$-letter capacity is also
provided which is tighter than the cut-set bound. The capacity is
the minimum of four $n$-letter mutual information terms for some joint $n$-letter
distribution. Therefore, each of these four $n$-letter terms can be
considered as an $n$-letter upper bound on the capacity.

The multiple access diamond channel is related to some other multiple access channel problems, such as sending arbitrarily correlated sources through the multiple access channel \cite{Cover:1980} (also see \cite{Ahlswede:1983}), and the multiple access channel with conferencing encoders \cite{Willems:1983, Bross:2008}.
The major difference between the multiple access diamond channel and the two multiple access channel problems above is the presence of a centralized
encoder in the multiple access diamond channel. This property enables one to construct a pair of correlated codes \cite{Ahlswede:1983,Traskov:2007} similar to the achievability of the general broadcast channel by Marton \cite{Marton:1979}. 

In this paper, we focus on the multiple access diamond channel where
the multiple access channel from the two relay nodes to the
destination node is Gaussian. We call this channel model \emph{the
Gaussian multiple access diamond channel}.
We first obtain an upper bound on the capacity via
single-letterization of one of the $n$-letter upper bounds in
\cite{Traskov:2007}. The main technique in the upper bound
derivation is bounding the correlation between the transmitted signals
 of the two relays via an auxiliary random variable. This
technique was used by Ozarow  in solving the Gaussian multiple
description problem \cite{Ozarow:1980}.

As for the lower bound, we first provide an achievable rate for the
general, i.e., not necessarily Gaussian, multiple access diamond
channel. The achievability scheme we propose is similar to
\cite{Ahlswede:1983,Traskov:2007}, except that the codebook is of a
superposition structure. The inner code, which contains part of the
message, is decoded by both relay nodes and serves as common data.
The outer code, which contains the remaining part of the message, is
a pair of correlated codewords and Codeword $k$ is delivered to Relay $k$, $k=1,2$.  The relay nodes send correlated codewords along
with the common data into the multiple access channel. This
achievable rate is then specialized  to the Gaussian multiple access
diamond channel. 

Finally, we characterize the sufficient and necessary conditions under which the proposed upper bound is strictly tighter than the cut-set bound. Furthermore, noting that the proposed upper and lower bounds take on
similar forms, we proceed to provide sufficient and necessary conditions under which our
upper and lower bounds meet. Thus, for a Gaussian multiple access
diamond channel that satisfies these conditions, we have found its
capacity.

The remainder of this paper is organized as follows. In Section II,
we provide the system model. In Section III, we derive an upper
bound on the capacity of the Gaussian multiple access diamond
channel. Achievable schemes and the corresponding rates are
described in Section IV. In Section V, we provide conditions under which
our upper and lower bounds meet and thus provide the capacity of
Gaussian multiple access diamond channels that satisfy these
conditions. Proofs are collected in Section VI, which are followed by
conclusions in Section VII.

\section{System Model}
\begin{figure}[t!]
\centering
\includegraphics[width=3.5in]{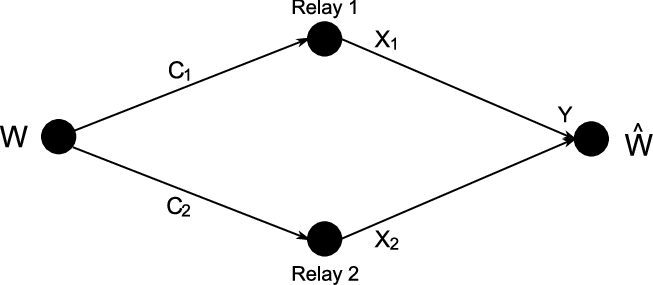}
\caption{The multiple access diamond channel.} \label{MAC_diamond}
\end{figure}

Consider a multiple access diamond channel, see Figure
\ref{MAC_diamond}. The capacity of the link from the source node to
Relay $k$ is $C_k$, for $k=1,2$. The channel between the two relay nodes
and the destination node is a multiple access channel with input
alphabets $(\mathcal{X}_1, \mathcal{X}_2)$, output alphabet
$\mathcal{Y}$ and a transition probability $p(y|x_1,x_2)$ defined on
$\mathcal{Y}\times\mathcal{X}_1\times\mathcal{X}_2$. Let $W$ be a
message that the source node would like to transmit to the
destination node. Assume that $W$ is uniformly distributed on
$\{1,2,\cdots,M\}$. An $(M, n, \epsilon_n)$ code consists of an
encoding function at the source node
\begin{align}
f^n&: \{1,2,\cdots,M\} \rightarrow \{1,2,\cdots,2^{nC_1}\} \times
\{1,2,\cdots, 2^{nC_2}\},\nonumber
\end{align}
two encoding functions at the relays
\begin{align}
f_k^n&: \{1,2,\cdots, 2^{nC_k}\} \rightarrow \mathcal{X}_k^n, \quad
k=1,2, \label{relay_encode}
\end{align}
and a decoding function at the destination node
\begin{align}
g^n: \mathcal{Y}^n \rightarrow \{1,2,\cdots,M\}. \nonumber
\end{align}
The average probability of error is defined as
\begin{align}
\epsilon_n=\sum_{w=1}^{M} \frac{1}{M} \textsf{Pr}[g^n(Y^n) \neq
w|W=w] \nonumber
\end{align}
Rate $R$ is said to be achievable if there exists a sequence of $(2^{nR},n, \epsilon_n)$ codes such that
$\epsilon_n \rightarrow 0$ as $n \rightarrow \infty$. The capacity of the
multiple access diamond channel is the supremum of all achievable
rates.

\begin{figure}[t!]
\centering
\includegraphics[width=3.5in]{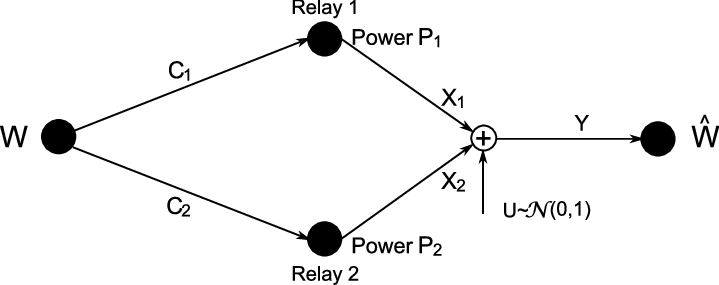}
\caption{The Gaussian multiple access diamond channel.} \label{dia}
\end{figure}

In this paper, we focus on the Gaussian multiple access diamond
channel, i.e., $\mathcal{X}_1=\mathcal{X}_2=\mathcal{Y}=\mathbb{R}$
and the channel between the two relay nodes to the destination node
is a Gaussian multiple access channel, see Figure \ref{dia}. The
received signal at the destination node is
\begin{align}
Y=X_1+X_2+U, \nonumber
\end{align}
where $X_1$ and $X_2$ are the input signals from Relay 1 and Relay
2, respectively, and $U$ is a zero-mean unit-variance Gaussian
random variable. It is assumed that $U$ is independent to
$(X_1,X_2)$. The encoding functions at the two relay nodes must
satisfy the average power constraints: for any codeword $x_k^n$ that Relay $k$
sends into the Gaussian multiple access channel, it satisfies
\begin{align}
\frac{1}{n}\sum_{i=1}^nx_{ki}^2&\le P_k, \quad k=1,2. \nonumber
\end{align}


We would like to characterize the capacity of the Gaussian multiple
access diamond channel in terms of the channel parameters $C_1, C_2,
P_1$ and $P_2$.

To simplify presentation, we define the following functions of
$\rho$ for $\rho \in [0,1]$:
\begin{align}
f_1(\rho) &\overset{\triangle}{=}C_1+\frac{1}{2}\log[1+(1-\rho^2)P_2], \nonumber\\
f_2(\rho) &\overset{\triangle}{=}C_2+\frac{1}{2}\log[1+(1-\rho^2)P_1], \nonumber\\
f_3(\rho) &\overset{\triangle}{=}C_1+C_2-\frac{1}{2}\log\left(\frac{1}{1-\rho^2}\right), \nonumber\\
f_4(\rho)
&\overset{\triangle}{=}\frac{1}{2}\log\left(1+P_1+P_2+2\rho\sqrt{P_1P_2}\right).
\nonumber
\end{align}
\section{An Upper Bound}
We first provide an upper bound on the capacity of the Gaussian
multiple access diamond channel.

\begin{Theo}\label{nece}
An upper bound on the capacity of the Gaussian multiple access diamond channel, denoted as $C_{\text{upper}}$, 
is
\begin{align}
C_{\text{upper}}=\max(T_1,T_2), \nonumber
\end{align}
where
\begin{align}
T_1&=\max_{0\le\rho\le\rho^*}\min\left\{f_1(\rho), f_2(\rho), f_3(\rho), f_4(\rho)\right\}, \label{T1}\\
T_2&=\max_{\rho^*\le\rho\le1}\min\left\{f_1(\rho), f_2(\rho),
f_3(0), f_4(\rho)\right\}, \nonumber
\end{align}
and
\begin{equation}
\rho^*=\sqrt{1+\frac{1}{4P_1P_2}}-\frac{1}{2\sqrt{P_1P_2}}.
\label{rhostar}
\end{equation}
%
\end{Theo}
The proof of Theorem \ref{nece} is in Section \ref{proof_nece}.

\noindent\emph{Remark}: 
%
We note that $T_1$ and $T_2$ essentially take the same form except the third term which is equal to $f_3(\rho)$ for $0\le\rho\le\rho^*$ in $T_1$ and equal to $f_3(0)$ for $\rho^*\le\rho\le1$ in $T_2$.
The cut-set bound for the Gaussian multiple access
diamond channel, denoted as $C_{\text{cut}}$, is
\begin{align}
C_{\text{cut}}=\max_{ 0 \leq \rho \leq 1} \min \{f_1(\rho), f_2(\rho), f_3(0),
f_4(\rho)\}. \label{ReviewCutSet}
\end{align}
Notice that the third term is $f_3(0)$ for $0 \leq \rho \leq 1$.
Hence, our upper bound is in general tighter than the cut-set bound. Our upper bound is strictly tighter than the cut-set bound when $C_{\text{upper}}$ takes the value of $f_3(\rho)$ for some $\rho$ satisfying $0\le\rho\le\rho^*$. This will be illustrated analytically in Theorem \ref{capacity} and also numerically by examples in Section \ref{sec_capacity}.

The converse result in Theorem \ref{nece} is a single-letterization
of one of the $n$-letter upper bounds in \cite{Traskov:2007}, which
is tighter than the cut-set bound.
In the cut-set bound, the cut through the two orthogonal links
yields $f_3(0)=C_1+C_2$, which implies that if it is achievable, the
signals through the two links should be independent. However, to
achieve a larger rate in the second stage, i.e., from the two relays
to the destination, the inputs of the multiple access channel should
be correlated, in which case $C_1+C_2$ is no longer achievable. To
obtain a tighter bound, it is essential to characterize the
correlation between the code pair $X_1^n$ and $X_2^n$,
i.e.,~$I(X_1^n;X_2^n)$, where $X_k^n$ is the length $n$ sequence transmitted by relay $k$, $k=1,2$. This is why in the upper bound in
\cite{Traskov:2007}, $C_1+C_2$ is replaced by
$C_1+C_2-\frac{1}{n}I(X_1^n;X_2^n)$. However it is in an $n$-letter
form and therefore uncomputable. 

It is desirable to ``single-letterize" the term
$I(X_1^n;X_2^n)$ for a pair of correlated codes, not only in our
problem of the multiple access diamond channel, but also in other
problems in multi-user information theory. This single-letterization problem remains open
in its general form. But for the Gaussian multiple description
problem, $I(X_1^n; X_2^n)$ is single-letterized in
\cite{Ozarow:1980} by introducing an auxiliary random variable and
applying the entropy power inequality \cite{Bergmans:1974}. Inspired
by \cite{Ozarow:1980}, we define the auxiliary random variables
$Z_i=Y_i+U'_i$, $i=1,2,\cdots,n$, where $Y_i$ is the output of the multiple access channel of the $i$-th channel use and $U'_i$ is a Gaussian random variable with zero mean and variance
$N$ and it is independent to everything else. By choosing the right value of $N$, we can obtain a tight upper bound on the term $I(X_1^n;X_2^n)$.
Such non-negative $N$ can be found
 as long as the correlation
between $X_1^n$ and $X_2^n$ is not too large. This is the reason why in
Theorem \ref{nece}, the upper bound $f_3(\rho)$ can only be enforced
for $\rho \leq \rho^*$. The details of the proof is provided in
Section \ref{proof_nece}.

\section{A Lower Bound}
In this section, we first provide a lower bound on the capacity of
the general, i.e., not necessarily Gaussian, multiple access diamond
channel. We then specialize the result to the Gaussian case and
obtain an achievable rate for the Gaussian multiple access diamond
channel. 
\begin{Theo}\label{achiev}
A lower bound on the capacity of the multiple access diamond channel
is
\begin{align}
\min\left\{
\begin{array}{l}
C_1+C_2-R_0-I(X_1;X_2|V)\\
C_1+I(X_2;Y|X_1,V)\\
C_2+I(X_1;Y|X_2,V)\\
I(X_1,X_2;Y)\\
I(X_1,X_2;Y|V)+R_0
\end{array}
\right\} \nonumber
\end{align}
for some $0\le R_0\le \min(C_1,C_2)$ and some joint distribution
$p(v,x_1,x_2)$.
\end{Theo}

The details of the proof is provided in Section \ref{proof_achiev}.
Here, we give a brief outline. The achievability scheme we use in
obtaining Theorem \ref{achiev} is the following: the source node
uses a codebook of the superposition structure, where $R_0$ is the
rate of the inner codebook. The source node splits its message into two
parts, with rates $R_0$ and $R-R_0$, respectively.
The source node encodes the first part of the message into the inner code and delivers the corresponding codeword index to both relays. The source node encodes the second part of the message into the outer code, which consists of a pair of correlated codewords.
The index of the first codeword of the pair is sent to Relay $1$, while
the index of the second codeword of the pair is sent to Relay $2$. Each relay finds the codeword from the codebook corresponding to its received indices, and
sends the codeword into the multiple access channel. An independent and concurrent version of Theorem \ref{achiev} is in \cite{Bidokhti:2014}.

From Theorem \ref{achiev}, we obtain an achievable rate for the
Gaussian multiple access diamond channel in the following Corollary.
\begin{Cor} \label{Gaussach}
A lower bound on the capacity of the Gaussian multiple access
diamond channel is
\begin{align}
\min\left\{
\begin{array}{l}
C_1+C_2-R_0-\frac{1}{2}\log\frac{1}{1-\rho^2}\\
C_1+\frac{1}{2}\log[1+(1-\rho^2)(1-\beta^2)P_2]\\
C_2+\frac{1}{2}\log[1+(1-\rho^2)(1-\alpha^2)P_1]\\
\frac{1}{2}\log[1+P_1+P_2+2\sqrt{P_1P_2}(\alpha\beta+\rho\sqrt{(1-\alpha^2)(1-\beta^2)})]\\
\frac{1}{2}\log[1+(1-\alpha^2)P_1+(1-\beta^2)P_2+2\rho\sqrt{(1-\alpha^2)(1-\beta^2)P_1P_2}]+R_0
\end{array}
\right\}  \label{Gauss_achrate}
\end{align}
for some $0\le R_0\le \min(C_1,C_2)$, $0\le\alpha,\beta\le 1$.
\end{Cor}
\begin{IEEEproof}
Consider the following configuration of $X_1,X_2$ and $V$:
\begin{align}
X_1&=\alpha\sqrt{P_1}V+X_1', \nonumber\\
X_2&=\beta\sqrt{P_2}V+X_2', \nonumber
\end{align}
where $V \sim\mathcal{N}(0,1)$, and $(X_1',X_2')$ are jointly Gaussian
with zero mean and covariance matrix
\begin{align}
\begin{bmatrix}
(1-\alpha^2)P_1&\rho\sqrt{(1-\alpha^2)(1-\beta^2)P_1P_2}\\
\rho\sqrt{(1-\alpha^2)(1-\beta^2)P_1P_2}&(1-\beta^2)P_2
\end{bmatrix}. \nonumber
\end{align}
Furthermore, $(X_1',X_2')$ and $V$ are independent. With this
selection of $X_1,X_2,V$, the achievable rate of
(\ref{Gauss_achrate}) follows from Theorem \ref{achiev} easily.
\end{IEEEproof}

In Corollary \ref{Gaussach}, if we set $R_0=\alpha=\beta=0$, we
obtain a smaller achievable rate:
\begin{Cor}\label{suff}
A lower bound of the capacity of the Gaussian multiple access
diamond channel, denoted as $C_{\text{lower}}$, is
\begin{align}
C_{\text{lower}}=\max_{0\le\rho\le1}\min\left\{f_1(\rho), f_2(\rho),
f_3(\rho), f_4(\rho)\right\}. \label{achach}
\end{align}
\end{Cor}

\emph{Remark}: The achievable rate in Corollary \ref{suff} is in general smaller than that in Corollary \ref{Gaussach}. More specifically, when $C_1$ and $C_2$ are sufficiently large, the multiple access channel in the second stage is the bottleneck of the network, and the optimal achievable scheme is sending fully correlated codewords into the multiple access channel, resulting in the achievable rate of $f_4(1)$. But the correlated codes without the superposition structure, used in Corollary \ref{suff}, can not support fully correlated codewords, because $f_4(1)>f_3(1)$, which means $C_{\text{lower}}<f_4(1)$. This shows that the superposition scheme in Corollary \ref{Gaussach} strictly outperforms the scheme without superposition structure in the case with large $C_1$ and $C_2$.

Though the lower bound in Corollary \ref{suff} is smaller than that of Corollary
\ref{Gaussach}, it takes on a similar form as the upper bound in
Theorem \ref{nece}. Thus, when the channel satisfy certain
conditions, we expect the upper bound in Theorem \ref{nece} and the
lower bound in Corollary \ref{suff} to coincide, yielding the
capacity. This will be discussed in the next section.

\section{Capacity} \label{sec_capacity}
Comparing the upper and lower bounds proposed in Theorem \ref{nece}
and Corollary \ref{suff}, we see that they take on similar forms.
More specifically, the four functions after the minimum in
(\ref{T1}) are exactly the same as that in (\ref{achach}).  Thus, it can be expected that for certain
parameters of the Gaussian multiple access diamond channel,
 $C_1, C_2, P_1$ and $P_2$, the upper and lower bounds meet providing us with the exact capacity of the channel.
We now proceed to give explicit conditions on the channel
parameters such that our upper and lower bounds meet.

\begin{figure}[h]
\centering
\includegraphics[width=4in]{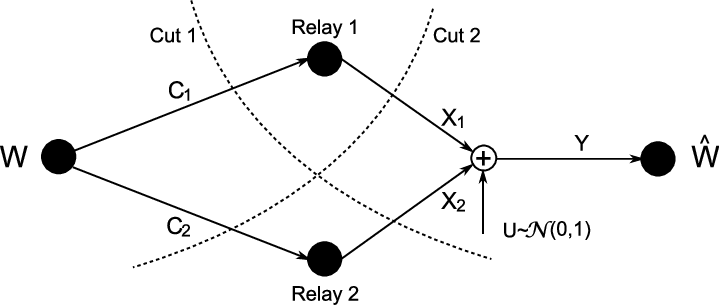}
\caption{The cross-cuts of the diamond channel.} \label{cut_fig}
\end{figure}

 First, we eliminate the trivial cases. If the channel is such that
\begin{align}
\min\left(C_1,C_2\right)\geq \frac{1}{2} \log
\left(1+P_1+P_2+2\sqrt{P_1P_2}\right), \nonumber
\end{align}
then the multiple access channel in the second stage is the
bottleneck of the whole network, and thus, the capacity of the
Gaussian multiple access diamond channel is equal to
\begin{align}
\frac{1}{2} \log \left(1+P_1+P_2 \break +2\sqrt{P_1P_2} \right).
\nonumber
\end{align}
On the other hand, if the channel is such that
\begin{align}
\min
\left(C_1+\frac{1}{2}\log(1+P_2),C_2+\frac{1}{2}\log(1+P_1),C_1+C_2
\right) \leq \frac{1}{2} \log (1+P_1+P_2), \nonumber
\end{align}
then the two separate links in the first stage, or one of the two
cross-cuts, i.e., Cut 1 or Cut 2 in Figure \ref{cut_fig}, is the
bottleneck of the whole network, and the capacity is equal to
\begin{align}
\min
\left(C_1+\frac{1}{2}\log(1+P_2),C_2+\frac{1}{2}\log(1+P_1),C_1+C_2
\right). \nonumber
\end{align}

Thus, we only need to focus on the nontrivial cases where
\begin{align}
\min\left(C_1,C_2\right)& < \frac{1}{2} \log
\left(1+P_1+P_2+2\sqrt{P_1P_2}\right), \textsf{ and }\label{cond1}\\
\frac{1}{2} \log (1+P_1+P_2) &< \min
\left(C_1+\frac{1}{2}\log(1+P_2),C_2+\frac{1}{2}\log(1+P_1),C_1+C_2
\right) \label{cond2}
\end{align}
are both satisfied.
In the next theorem, we provide sufficient and necessary conditions on the channel parameters for the nontrivial cases described in (\ref{cond1}) and (\ref{cond2}), 
such that the upper and lower bounds proposed in Theorem \ref{nece}
and Corollary \ref{suff} meet, yielding the exact capacity, and also the conditions, under which the cut-set bound is strictly larger than the proposed upper bound. 
\begin{Theo} \label{capacity}
Consider a
 Gaussian multiple access diamond channel that
satisfies (\ref{cond1}) and (\ref{cond2}). Define $f_5(\rho)=\min
(f_1(\rho), f_2(\rho)), \forall \rho \in [0,1]$. Let $\bar{\rho}_k$
denote the positive root of the equation $f_k(\rho)=f_4(\rho)$  that is in
$[0,1]$, $k=3,5$. Recall that $\rho^*$ is defined in Theorem 1.
\begin{enumerate}
\item 
The sufficient and necessary condition for $C_{\text{lower}}=C_{\text{upper}}$ to hold is
\begin{align}
\bar{\rho}_5 \leq \bar{\rho}_3 \quad \text{ or } \quad
f_4(\bar{\rho}_3)  \geq f_5(\rho^*),\nonumber
\end{align}
and in this case, the capacity is $f_4(\bar{\rho}_3)$. 
\item The sufficient and necessary condition for $C_{\text{upper}}<C_{\text{cut}}$ to hold is
\begin{align}
\bar{\rho}_3 < \bar{\rho}_5 \quad \text{ and } \quad \rho^* \geq \bar{\rho}_5 \quad \text{ and } \quad f_3(0) > f_5(\rho^*)
\end{align}
\item The sufficient and necessary condition for $C_{\text{lower}}=C_{\text{upper}}<C_{\text{cut}}$ to hold is
\begin{align}
\bar{\rho}_3 < \bar{\rho}_5 \quad \text{ and } \quad f_4(\bar{\rho}_3)  \geq f_5(\rho^*) \label{cww_con7}
\end{align}
and in this case, the capacity is $f_4(\bar{\rho}_3)$. 
\end{enumerate}
\end{Theo}

The proof of Theorem \ref{capacity} is provided in Section
\ref{proof_capacity}. Theorem \ref{capacity} illustrates the tightness of the novel upper bound on capacity proposed in Theorem 1, in that for some channel parameters, it is achievable and strictly smaller than the cut-set bound. To the best of our knowledge, Theorem 3 demonstrates the first time where the capacity of certain Gaussian multiple access diamond channels has been characterized when the cut-set bound is not tight. 

To show that there indeed exist Gaussian multiple access diamond
channels that
 satisfy the condition (\ref{cww_con7}) stated in Theorem \ref{capacity}, we give the following examples:
\begin{enumerate}
\item $C_1=1.8, C_2=2, P_1=10, P_2=15$, which means
$\rho^*=0.9600, \bar{\rho}_3=0.8696,
\bar{\rho}_5=0.8947, f_4(\bar{\rho}_3)=2.7819, f_5(\rho^*)=2.3608$.
Thus,
(\ref{cww_con7}) is satisfied and the
capacity is $f_4(\bar{\rho}_3)=2.7819$ bits/channel use.
\item $C_1=2, C_2=1.2, P_1=30, P_2=20,$ which means
$\rho^*=0.9798, \bar{\rho}_3=0.4029,
\bar{\rho}_5=0.7027, f_4(\bar{\rho}_3)=3.0722, f_5(\rho^*)=1.7688$.
Thus,
(\ref{cww_con7}) is satisfied and the
capacity is $f_4(\bar{\rho}_3)=3.0722$ bits/channel use.
\end{enumerate}

Since the conditions provided in Theorem \ref{capacity} is rather involved, we provide some intuition as to when our proposed upper bound in Theorem \ref{nece}
and the lower bound in Corollary \ref{suff} meet. As noted after the statement of Theorem \ref{nece}, the upper bound $f_3(\rho)$ can only be enforced
for $\rho \leq \rho^*$. We also observe that $\rho^*$  monotonically increases with respect to $P_1$ and $P_2$. Therefore, the proposed upper bound is tight with large $P_1$ and/or $P_2$.
From the perspective of achievability, as discussed in the remark after Corollary \ref{suff}, when $C_1$ and $C_2$ are large, the superposition scheme in Corollary \ref{Gaussach} outperforms the scheme without the superposition structure in Corollary \ref{suff}, i.e., the rate characterized by Corollary \ref{suff} is not tight in this case. From the above discussions, we can see that the proposed upper bound in Theorem \ref{nece} and the lower bound in Corollary \ref{suff} meet when $P_1$ and $P_2$ are large and/or $C_1$ and $C_2$ are small. This is also supported by our numerical results discussed next. 
\begin{figure}
\centering
\includegraphics[width=7in]{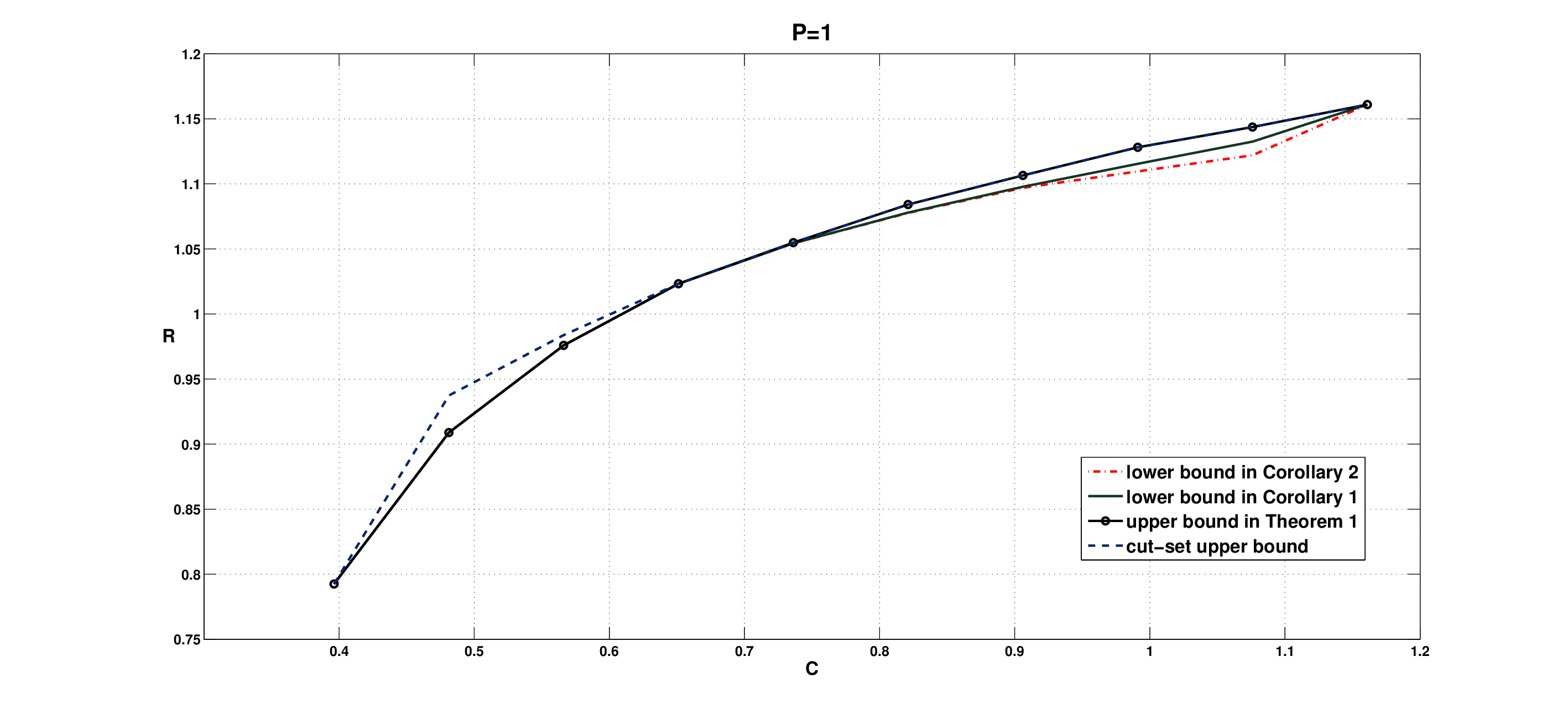}
\caption{Comparison of upper and lower bounds with $P=1$.}
\label{P1}
\end{figure}
\begin{figure}
\centering
\includegraphics[width=7in]{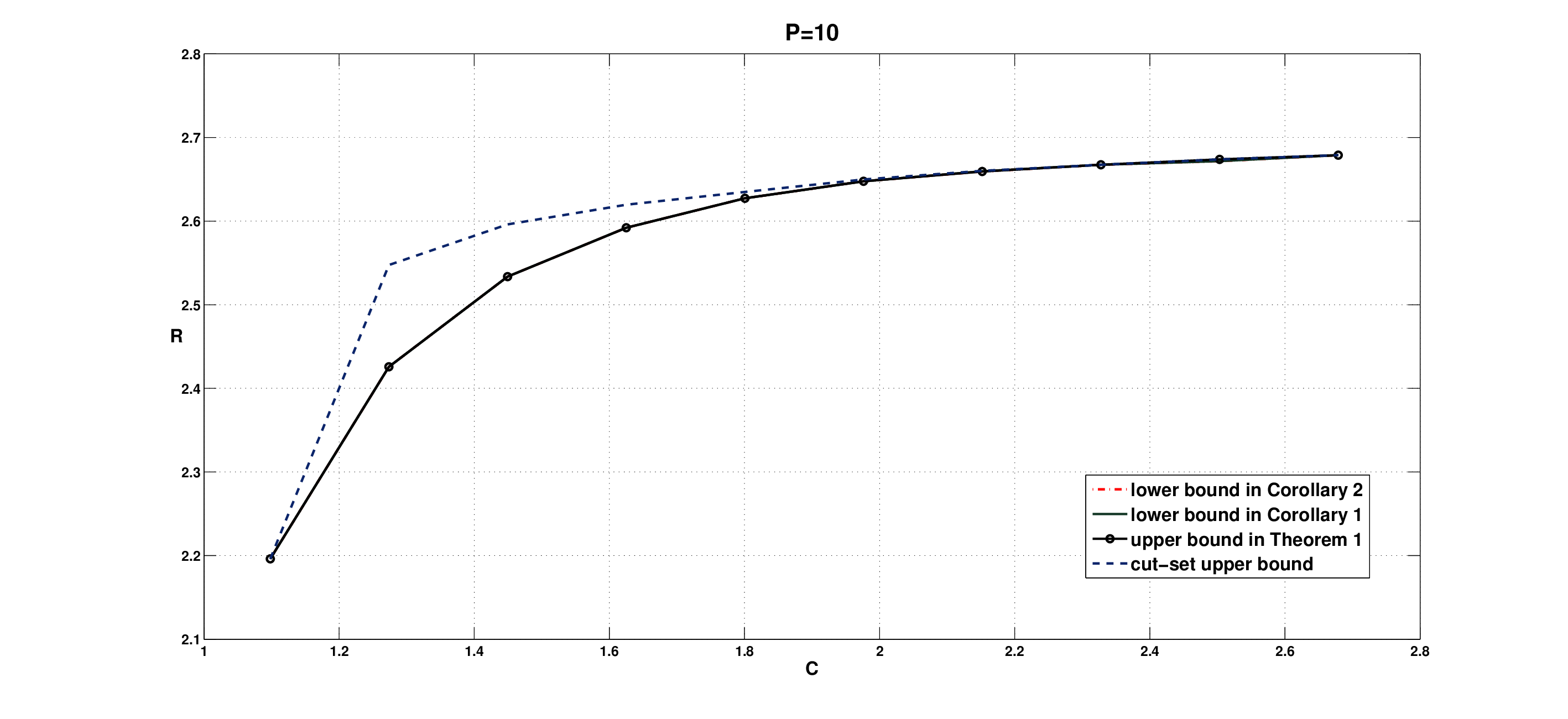}
\caption{Comparison of upper and lower bounds with $P=10$.}
\label{P5}
\end{figure}

We plot the upper and lower bounds in Theorem \ref{nece},
Corollary \ref{Gaussach} and Corollary \ref{suff} and depict them in
Figures \ref{P1} and \ref{P5} for the symmetric cases of
$C_1=C_2\overset{\triangle}{=}C, P_1=P_2\overset{\triangle}{=}P=1$
and $C_1=C_2\overset{\triangle}{=}C,
P_1=P_2\overset{\triangle}{=}P=10$, respectively. The cut-set bound
is also plotted to show the improvement of our upper bound over the
cut set bound, which was the best known upper bound on the capacity.
As can be seen, for small $C$, the proposed upper bound is strictly
smaller than the cut-set bound. Also, the gap between our lower and
upper bounds is rather small, especially when $C$ is relatively
small and/or $P$ is relatively large.

\section{Proofs}
\subsection{Proof of Theorem \ref{nece}} \label{proof_nece}
In order to make the proof easier to follow, we split the entire proof into 4 steps, consisting of Lemmas 1 to 4. 

Starting from Fano's inequality, any achievable rate $R$ must satisfy
\begin{align}
nR& =H(W) \nonumber\\
&=I(W;Y^n)+H(W|Y^n)\nonumber\\
& \leq I(X_1^n, X_2^n;Y^n)+H(W|Y^n) \label{Markov}\\
& \leq I(X_1^n, X_2^n;Y^n)+n \epsilon_n. \label{Fano02}
\end{align}
where 
(\ref{Markov}) is because of the Markov
chain $W \rightarrow (X_1^n, X_2^n) \rightarrow Y^n$.

Towards proving that $R \leq \min (f_1(\rho), f_2(\rho))$ for some $\rho \in [0,1]$, we first show the following Lemma.

\begin{Lem} \label{Lemma1}
 There exists $\rho_a \in [0,1]$ and $\rho_b \in [0,1]$ such that 
\begin{align}
R \leq \frac{1}{2}\log\left[(1-\rho_a^2)P_1+1\right]+C_2+
\epsilon_n, \\
R \leq \frac{1}{2}\log\left[(1-\rho_b^2)P_2+1\right]+C_1+
\epsilon_n. 
\end{align}
\end{Lem}

\begin{IEEEproof}
Following from (\ref{Fano02}), we have
\begin{align}
n R
&\le I(X_1^n, X_2^n;Y^n)+n \epsilon_n\label{fano2again}\\
&= I(X_1^n;Y^n|X_2^n)+I(X_2^n;Y^n)+n \epsilon_n\nonumber\\
&\le I(X_1^n;Y^n|X_2^n)+H(X_2^n)+n \epsilon_n\nonumber\\
&\le I(X_1^n;Y^n|X_2^n)+n C_2+n \epsilon_n \label{lateradd01} \\
&\le \sum_{i=1}^n I(X_{1i};Y_i|X_{2i})+n C_2+  n \epsilon_n \label{lateradd02}\\
&\le\sum_{i=1}^n\frac{1}{2}\log\left[(1-\rho_i^2)P_{1i}+1\right]+n
C_2+  n \epsilon_n, \label{Gaussmax}
\end{align}
(\ref{lateradd01}) is because without loss of generality, we only
consider deterministic encoders, thus from (\ref{relay_encode}) we have $H(X_k^n) \leq n C_k$, $k=1,2$; (\ref{lateradd02}) follows
from the memoryless nature of the channel $p(y|x_1,x_2)$; 
(\ref{Gaussmax}) follows because we have defined $P_{ki}
\overset{\triangle}{=} E[X_{ki}^2], k=1,2$ and
$\rho_i=\frac{E[X_{1i}X_{2i}]}{\sqrt{P_{1i}P_{2i}}}$, and used the
fact that given power constraint, the Gaussian distribution
maximizes the differential entropy \cite{Cover:1991}. 

Since the inputs from Relay 1 must satisfy the average power
constraint $P_1$, we have
\begin{equation}
0\le\frac{1}{n}\sum_{i=1}^n\rho_i^2 P_{1i}\le
\frac{1}{n}\sum_{i=1}^nP_{1i} \leq P_1. \nonumber
\end{equation}
Therefore, there exists a $\rho_a \in [0,1]$  such that
\begin{equation}
\rho_a^2 P_1=\frac{1}{n}\sum_{i=1}^n\rho_i^2 P_{1i}. \nonumber
\end{equation}
Due to the concavity of the logarithm function, we have
\begin{align}
R &\le\frac{1}{n}\sum_{i=1}^n\frac{1}{2}\log[(1-\rho_i^2)P_{1i}+1]+C_2+ \epsilon_n\nonumber\\
&\le\frac{1}{2}\log\left(\frac{1}{n}\sum_{i=1}^n\left[(1-\rho_i^2)P_{1i}+1\right]\right)+C_2+ \epsilon_n\nonumber\\
&=\frac{1}{2}\log\left(\frac{1}{n}\sum_{i=1}^n P_{1i}
-\frac{1}{n}\sum_{i=1}^n \rho_i^2 P_{1i}
 +1 \right) +C_2+ \epsilon_n\nonumber\\
& \leq \frac{1}{2}\log\left[(1-\rho_a^2)P_1+1\right]+C_2+
\epsilon_n. \label{concave}
\end{align}
Similarly, there exists a $\rho_b \in [0,1]$  such that
\begin{equation}
\rho_b^2 P_2=\frac{1}{n}\sum_{i=1}^n \rho_i^2 P_{2i}, \nonumber
\end{equation}
and we have
\begin{align}
R&\le\frac{1}{2}\log\left[(1-\rho_b^2)P_2+1\right]+C_1+ \epsilon_n. \label{sym01}
\end{align}
This completes the proof of Lemma \ref{Lemma1}. 
\end{IEEEproof}

The following lemma shows the relationship between the achievable rate $R$ and the n-letter entropy $h(Y^n)$.  

\begin{Lem} \label{Lemma2}
\begin{align} 
 h(Y^n)-\frac{n}{2} \log (2 \pi e) \leq nR \leq  h(Y^n)-\frac{n}{2} \log (2 \pi e)+n \epsilon_n. \nonumber
\end{align}
\end{Lem}

\begin{IEEEproof}
From (\ref{Fano02}), we have
\begin{align}
n R& \leq h(Y^n)-h(Y^n|X_1^n, X_2^n)+n \epsilon_n \nonumber\\
&= h(Y^n)-\frac{n}{2} \log (2 \pi e)+n \epsilon_n. \nonumber
\end{align}
On the other hand, we also have
\begin{align}
n R&=H(W) \nonumber\\
& \geq H(X_1^n, X_2^n) \label{function}\\
& \geq I(X_1^n,X_2^n;Y^n)\nonumber \\
& =h(Y^n)-\frac{n}{2} \log (2 \pi e), \label{lowerR}
\end{align}
where (\ref{function}) is because 
$(X_1^n, X_2^n)$ is a deterministic function of $W$. This completes the proof of Lemma \ref{Lemma2}. 
\end{IEEEproof}

From Lemma \ref{Lemma2}, we see that to characterize the achievable rate $R$, we need to characterize $h(Y^n)$.
Towards this end, let us define $\rho \in [0,1]$, which is a function of $h(Y^n)$ as
follows: If
\begin{align}
\frac{1}{n}h(Y^n) \leq \frac{1}{2}\log (2 \pi e)(1+P_1+P_2),
\label{triv}
\end{align}
then  $\rho=0$; otherwise, $\rho$ is such that
\begin{align}
\frac{1}{n}h(Y^n)=\frac{1}{2}\log (2 \pi
e)(1+P_1+P_2+2\rho\sqrt{P_{1}P_{2}}). \label{bound0}
\end{align}

\begin{Lem}  \label{Lemma3}
$\rho$ as defined above satisfies $\rho \leq \min(\rho_a, \rho_b)$. 
\end{Lem}

\begin{IEEEproof}
We upper bound $h(Y^n)$ as
\begin{align}
\frac{1}{n}h(Y^n) &\leq \frac{1}{n}\sum_{i=1}^n h(Y_i) \nonumber\\
&\leq \frac{1}{n}\sum_{i=1}^n\frac{1}{2}\log (2 \pi
e)\left(P_{1i}+P_{2i}+2\rho_i\sqrt{P_{1i}P_{2i}} +1\right),
\label{define_i}\\
&\le\frac{1}{2}\log (2 \pi e)
\left(\frac{1}{n}\sum_{i=1}^n\left[P_{1i}+P_{2i}
+2|\rho_i|\sqrt{P_{1i}P_{2i}}+1\right]\right)\label{after01}\\
& \leq \frac{1}{2}\log (2\pi
e)\left(P_1+P_2+\frac{1}{n}\sum_{i=1}^n2\sqrt{\rho_i^2P_{1i}P_{2i}}+1\right) \label{after02}.
\end{align}
where (\ref{define_i}) follows from the same reason as (\ref{Gaussmax}); (\ref{after01}) follows from the concavity of the logarithmic function; (\ref{after02}) follows from the same argument as (\ref{concave}). 
From Cauchy-Schwarz inequality, we have
\begin{align}
\frac{1}{n}\sum_{i=1}^n\sqrt{\rho_i^2P_{1i}P_{2i}}&\le
\sqrt{\left(\frac{1}{n}\sum_{i=1}^n\rho_i^2P_{1i}\right)\left(\frac{1}{n}\sum_{i=1}^nP_{2i}\right)}\leq
\sqrt{\rho_a^2P_1P_2}. \nonumber
\end{align}
Thus, we have
\begin{align}
\frac{1}{n}h(Y^n)\le\frac{1}{2}\log (2 \pi
e)\left(P_1+P_2+2\rho_a\sqrt{P_{1}P_{2}}+1\right). \label{sat_a}
\end{align}
By symmetry, we also have
\begin{align}
\frac{1}{n}h(Y^n) &\le \frac{1}{2}\log (2 \pi
e)\left(P_1+P_2+2\rho_b\sqrt{P_{1}P_{2}}+1\right). \label{sat_b}
\end{align}
From (\ref{sat_a}), (\ref{sat_b}) and the definition of $\rho$, we have that $\rho \leq \min(\rho_a, \rho_b)$ which completes the proof of Lemma \ref{Lemma3}.
\end{IEEEproof} 

%

As $n \rightarrow \infty$, Lemma \ref{Lemma1} together with Lemma \ref{Lemma3} means that $R \leq \min (f_1(\rho), f_2(\rho))$, where $\rho$ is defined right before Lemma \ref{Lemma3}. Lemma \ref{Lemma2} together with the definition of $\rho$ means that $R \leq f_4(\rho)$. Thus, we have shown that $R \leq \min (f_1(\rho), f_2(\rho), f_4(\rho))$ and it remains to show that if  $\rho$ further satisfies $0 \leq \rho\le \rho^*$, we have $R \leq f_3(\rho)$. 

If $\rho=0$, from the cut-set bound, we have $R \leq C_1+C_2 = f_3(0)$. As for the case of $0 < \rho\le \rho^*$, using Ozarow's idea in \cite{Ozarow:1980}, we have the following lemma. 

\begin{Lem} \label{Lemma4}
If $\rho$ further satisfies $0 < \rho\le \rho^*$, we have
\begin{align}
2R \le
\frac{1}{2}\log\left(1+P_1+P_2+2\rho\sqrt{P_1P_2}\right)+C_1+C_2-\frac{1}{2}\log\left(\frac{1}{1-\rho^2}\right)+2
\epsilon_n. \label{boundC4}
\end{align}
\end{Lem}

\begin{IEEEproof}
If $\rho$ satisfies $0 < \rho\le \rho^*$, which is equivalent
to $\sqrt{P_1 P_2} \left(\frac{1}{\rho}-\rho \right)-1\ge0$, we
define additional random variables
\begin{align}
Z_i=Y_i+U'_i,\qquad i=1,\dots,n, \nonumber
\end{align}
where $U'^n$ is an i.i.d. Gaussian sequence with mean zero and
variance
\begin{align}
N=\sqrt{P_1 P_2}\left (\frac{1}{\rho}-\rho\right)-1, \label{defineW}
\end{align}
and is independent to everything else. We have
\begin{align}
2nR &  \leq 2I(X_1^n,X_2^n;Y^n)+2 n \epsilon_n \label{fano2againagain}\\
&\le I(X_1^n,X_2^n;Y^n)+H(X_1^n,X_2^n)+2n \epsilon_n \nonumber\\
&=I(X_1^n,X_2^n;Y^n)+H(X_1^n)+H(X_2^n)-I(X_1^n;X_2^n)+2n \epsilon_n \nonumber\\
&\leq I(X_1^n,X_2^n;Y^n)+n C_1+n C_2-I(X_1^n;X_2^n)+2n \epsilon_n, \label{01}
\end{align}
where (\ref{fano2againagain}) is the same as (\ref{Fano02}), and (\ref{01}) follows from the same reasoning as (\ref{lateradd01}).
Note that
\begin{align}
I(X_1^n;X_2^n)
&=I(X_1^n;Z^n)-I(X_1^n;Z^n|X_2^n)+I(X_1^n;X_2^n|Z^n)\nonumber\\
&\ge I(X_1^n;Z^n)-I(X_1^n;Z^n|X_2^n)\nonumber\\
&= I(X_1^n,X_2^n;Z^n)-I(X_2^n;Z^n|X_1^n)-I(X_1^n;Z^n|X_2^n).
\label{02}
\end{align}
We further have
\begin{align}
I(X_1^n;Z^n|X_2^n)&\le \sum_{i=1}^n\frac{1}{2}\log\frac{(1-\rho_i^2)P_{1i}+1+N}{1+N} \label{useGaussmax}\\
&\le\frac{n}{2}\log\frac{(1-\rho^2)P_{1}+1+N}{1+N},
\label{useconcave}
\end{align}
where (\ref{useGaussmax}) follows by similar arguments as
(\ref{Gaussmax}), and (\ref{useconcave}) follows by using similar arguments
as (\ref{concave}) and the result of Lemma \ref{Lemma3}. Similarly, we have
\begin{align}
I(X_2^n;Z^n|X_1^n)
&\le\frac{n}{2}\log\frac{(1-\rho^2)P_{2}+1+N}{1+N}. \label{03}
\end{align}
We also have
\begin{align}
I(X_1^n,X_2^n;Z^n)&=h(Z^n)-h(Z^n|X_1^n,X_2^n)\nonumber\\
&= h(Z^n)-\sum_{i=1}^n\frac{1}{2}\log (2\pi e) (1+N).\nonumber
\end{align}
From entropy power inequality (EPI) \cite[Lemma I]{Bergmans:1974},
we have
\begin{align}
h(Z^n)&\ge
\frac{n}{2}\log\left[2^{\left(\frac{2}{n}h(Y^n)\right)}+2\pi e N
\right]. \nonumber
\end{align}
Therefore,
\begin{align}
h(Z^n)-h(Y^n)
&\ge\frac{n}{2}\log\left[1+\frac{2\pi e N}{2^{\left(\frac{2}{n}h(Y^n)\right)}}\right]\nonumber\\
&=\frac{n}{2}\log\left[1+\frac{N}{ P_{1}+P_{2}+2\rho\sqrt{P_{1}P_{2}}+1}\right] \label{exact}\\
&=\frac{n}{2}\log\frac{P_{1}+P_{2}+2\rho\sqrt{P_{1}P_{2}}+1+N}{
P_{1}+P_{2}+2\rho\sqrt{P_{1}P_{2}}+1}, \nonumber
\end{align}
where (\ref{exact}) follows from (\ref{bound0}). Thus,
\begin{align}
I&(X_1^n,X_2^n;Y^n)-I(X_1^n,X_2^n;Z^n) \le\frac{n}{2}\log\frac{
(N+1)(P_{1}+P_{2}+2\rho\sqrt{P_{1}P_{2}}+1)}{P_{1}+P_{2}+2\rho\sqrt{P_{1}P_{2}}+1+N}.
\label{04}
\end{align}
Using (\ref{01}), (\ref{02}), (\ref{useconcave}), (\ref{03}) and
(\ref{04}), we have
\begin{align}
2nR &\le\frac{n}{2}\log(P_{1}+P_{2}+2\rho\sqrt{P_{1}P_{2}}+1)+n
C_1+n
C_2 \nonumber\\
&\hspace{0.1in}
-\frac{n}{2}\log\frac{(P_{1}+P_{2}+2\rho\sqrt{P_{1}P_{2}}+1+N)(1+N)}{((1-
\rho^2)P_{1}+1+N)((1-\rho^2)P_{2}+1+N)}+2 n \epsilon_n. \nonumber
\end{align}
Plugging in $N$ defined in (\ref{defineW}), we have proved (\ref{boundC4}), which completes the proof of Lemma \ref{Lemma4}.
\end{IEEEproof}

Hence, for the case of $0 \leq \rho \leq \rho^*$, letting $n
\rightarrow \infty$, from Lemma \ref{Lemma2} and Lemma \ref{Lemma4}, we have proved $R \leq f_3(\rho)$.
As for the case where $\rho^* < \rho \leq 1$, though the result of Lemma \ref{Lemma4} no longer holds, from the cut-set bound, we always have $R \leq C_1+C_2$, which means $R \leq f_3(0)$.

For all cases of $\rho \in [0,1]$, we have proved that the
achievable rate satisfies  either 
\begin{align}
R \leq \max_{0\le\rho\le\rho^*}\min\left\{f_1(\rho), f_2(\rho), f_3(\rho), f_4(\rho)\right\}
\end{align} 
or 
\begin{align}
R \leq \max_{\rho^*\le\rho\le1}\min\left\{f_1(\rho), f_2(\rho),
f_3(0), f_4(\rho)\right\},
\end{align}
and
thus, Theorem 1 is proved.

\subsection{Proof of Theorem \ref{achiev}} \label{proof_achiev}
For a given distribution $p(v,x_1,x_2)$, consider a rate tuple
$(R_0, R_1, R_2, r_1, r_2)$ such that
\begin{align}
r_1+r_2 & \geq I(X_1;X_2|V)+\delta, \label{Rconstraint01}\\
R_1&\le I(X_1;Y,X_2|V),\label{Rconstraint03}\\
R_2&\le I(X_2;Y,X_1|V), \label{Rconstraint02}\\
R_1+R_2&\le I(X_1,X_2;Y|V)+I(X_1;X_2|V), \label{Rconstraint05}\\
R_0+R_1+R_2&\le I(X_1,X_2;Y)+I(X_1;X_2|V),\label{Rconstraint04}\\
R_0+R_1&\le C_1, \label{Rconstrainta}\\
R_0+R_2&\le C_2, \label{Rconstraintb}\\
0 \leq r_1 &\leq R_1, \label{Rconstraintc}\\
0 \leq r_2 &\leq R_2, \label{Rconstraintd}
\end{align}
for any $\delta>0$. We will show that rate $R$ defined as 
\begin{align}
R=R_0+R_1-r_1+R_2-r_2, \label{Rconstrainte}
\end{align}
is achievable.

%
%
%

\noindent\textbf{Codebook generation}: 
First randomly generate
$2^{nR_0}$ many $v^n$ sequences according to $p(v)$ and index them
as $v^n(1),\dots,v^n(2^{nR_0})$. The $v^n$ sequences constitute
the inner codebook. 


Conditioned on $v^n(i)$,
$i=1,\dots,2^{nR_0}$, for each $j=1,2,\dots,2^{n(R_1-r_1)}$, generate a subcodebook $C_1(i,j)$ consisting of $2^{nr_1}$ many $x_1^n$ sequence in a conditionally i.i.d. fashion
according to $p(x_1|v)$. We index the codeword sequences in the subcodebook $C_1(i,j)$ as $x_1^n(i,l_1)$ for $l_1=(j-1)2^{nr_1}+1,\dots,j2^{nr_1}$. 
Similarly, conditioned on $v^n(i)$,
$i=1,\dots,2^{nR_0}$, for each $k=1,2,\dots,2^{n(R_2-r_2)}$, generate a subcodebook $C_2(i,k)$ consisting of $2^{nr_2}$ many $x_2^n$ sequences in a conditionally i.i.d. fashion according to $p(x_2|v)$. We index the codeword sequences in the subcodebook $C_2(i,k)$ as $x_2^n(i,l_2)$ for $l_2=(k-1)2^{nr_2}+1,\dots,k2^{nr_2}$.

\begin{figure}
\centering
\includegraphics[width=6in]{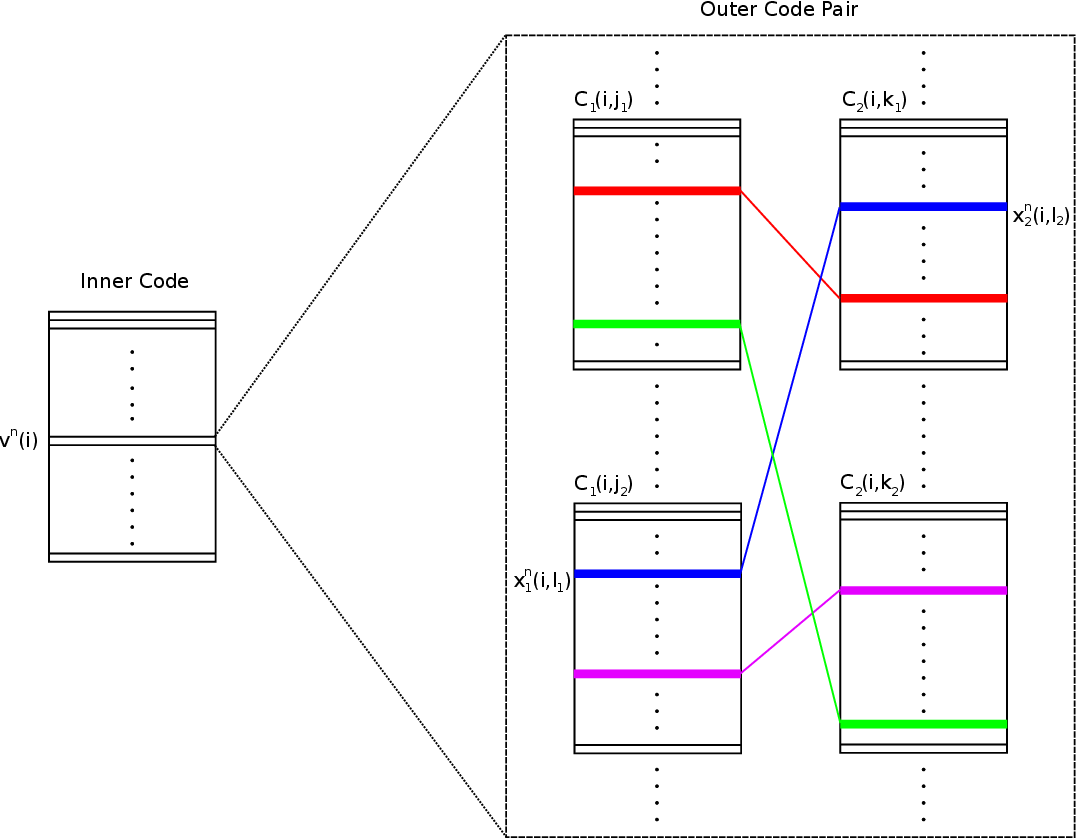}
\caption{The correlated codes with superposition structure.} \label{sup}
\end{figure}

For each subcodebook pair $(C_1(i,j),C_2(i,k))$, $i=1,\dots,2^{nR_0}$, $j=1,2,\dots,2^{n(R_1-r_1)}$ and $k=1,2,\dots,  2^{n(R_2-r_2)}$, find a pair of codewords $(x_1^n,x_2^n)$ such that $x_1^n\in C_1(i,j),x_2^n\in C_2(i,k)$ and
$(x_1^n,x_2^n)\in\mathcal{T}_{[X_1X_2|V]}^n(v^n(i))$,
where $\mathcal{T}_{[X_1X_2|V]}^n(v^n(i))$ is the conditional
typical set according to the distribution $p(x_1,x_2|v)$
\cite{Csiszar:1981}. If there are multiple such codeword pairs, then pick one pair randomly. If there is no such pair, we randomly choose a pair of $(x_1^n,x_2^n)$ from the subcodebook pair $(C_1(i,j),C_2(i,k))$. We denote the picked codeword pair, say $(x_1^n(i,l_1),x_2^n(i,l_2))$, as the $(i,j,k)$-th codeword pair. We also define the $i$-th correlated codebook $\mathcal{C}(i)$, which contains all the $(i,j,k)$-th codeword pair for $j=1,2,\dots,2^{n(R_1-r_1)}$ and $k=1,2,\dots,2^{n(R_2-r_2)}$. We illustrate the structure of inner code and the outer code graphically in Fig. \ref{sup}. 

\noindent\textbf{Encoding}: We split the message $W$, which is uniformly distributed on $\{1,2,\dots,2^{nR}\}$,
into $(W_a,W_b,W_c)$, where $W_a$, $W_b$ and $W_c$ are uniformly distributed on
$\{1,2,\dots,2^{nR_0}\}\break$, $\{1,2,\dots,2^{n(R_1-r_1)}\}$ and
$\{1,2,\dots,2^{n(R_2-r_2)}\}$, respectively. When $(W_a,W_b,W_c)=(i,j,k)$, we select the $(i,j,k)$-th
codeword pair from the subcodebook pair $(C_1(i,j),C_2(i,k))$, i.e., $(x_1^n(i,l_1),$ $x_2^n(i,l_2))$.
The transmitter sends index $(i,l_1)$ to
Relay $1$ and index $(i,l_2)$ to Relay $2$. The relays can correctly receive the indices due to $R_0+R_1\le C_1$ in (\ref{Rconstrainta}) and $R_0+R_2\le C_2$ in (\ref{Rconstraintb}). Upon receiving the index
$(i,l)$, Relay $1$ sends the codeword $x_1^n(i,l_1)$ into the multiple access channel and
similarly, upon receiving the index $(i,l_2)$, Relay $2$ sends the
codeword $x_2^n(i,l_2)$ into the multiple access channel.

\noindent\textbf{Decoding}: After receiving $y^n$, 
 if there exists
a unique codeword pair $(x_1^n(i,l_1), x_2^n(i,l_2))$ which is the $(i,j,k)$-th codeword pair from the 
subcodebook pair $(C_1(i,j),C_2(i,k))$, such that
\begin{align}
(v^n(i),x_1^n(i,l_1), x_2^n(i,l_2),y^n)\in\mathcal{T}_{[VX_1X_2Y]}^n, \nonumber
\end{align}
where $\mathcal{T}_{[VX_1X_2Y]}^n$ is the typical set as defined in
\cite{Csiszar:1981} according to $p(v,x_1,x_2,y)$, then the receiver
declares $(W_a,W_b,W_c)=(i,j,k)$; otherwise, the receiver declares an
error.

\noindent\textbf{Probability of Error}: Due to symmetry, the average
probability of error is equivalent to the probability of error for
an arbitrary message $w\in\{1,\dots,2^{nR}\}$. Hence, without loss of generality, we assume $W=w$, and the average probability of error satisfies
\begin{align}
\textsf{Pr}[E]&\triangleq \textsf{Pr} [g^n(Y^n)\ne w|W=w],
\qquad\qquad w=1,\dots,2^{nR}, \nonumber
\end{align}
For $W=w$, we denote $(W_a,W_b,W_c)=(i,j,k)$, and the corresponding codeword pair is denoted as
$(x_1^n(i,l_1),  x_2^n(i,l_2))$. An error occurs if one of the following error events happen.
\begin{enumerate}
\item $E_0$: there does not exist a pair of codewords $(x_1^n,x_2^n)$ such that $x_1^n\in C_1(i,j),x_2^n\in C_2(i,k)$ and
$(x_1^n,x_2^n)\in\mathcal{T}_{[X_1X_2|V]}^n(v^n(i))$.
\item $E_1$: $\left(v^n(i),x_1^n(i,l_1),x_2^n(i,l_2), Y^n\right)\not\in\mathcal{T}_{[VX_1X_2Y]}^n$.
\item $E_2$: There exist other codewords jointly typical with $Y^n$, which includes
\begin{enumerate}
\item $E_{21}$: there exists $l_1'\ne l_1$ such that $(x_1^n(i,l_1'),x_2^n(i,l_2)) \in \mathcal{C}(i)$ and
$(v^n(i),x_1^n(i,l_1'),$ $x_2^n(i,l_2),Y^n)\in\mathcal{T}_{[VX_1X_2Y]}^n$,
\item  $E_{22}$: there exists $l_2'\ne l_2$ such that $(x_1^n(i,l_1),x_2^n(i,l_2')) \in \mathcal{C}(i)$ and
$(v^n(i),x_1^n(i,l),$ $x_2^n(i,l_2'),Y^n)\in\mathcal{T}_{[VX_1X_2Y]}^n$;
\item  $E_{23}$: there exists $l_1'\ne l_1,l_2'\ne l_2$ such that $(x_1^n(i,l_1'),x_2^n(i,l_2')) \in \mathcal{C}(i)$ and
$(v^n(i),$ $x_1^n(i,l_1'),x_2^n(i,l_2'),Y^n)\in\mathcal{T}_{[VX_1X_2Y]}^n$;
\item  $E_{24}$: there exists $i' \ne i$, $l_1'$, $l_2'$ such that  $(x_1^n(i',l_1'),x_2^n(i',l_2')) \in \mathcal{C}(i')$ and $(v^n(i'), x_1^n(i',l_1'),$ $x_2^n(i',l_2'),Y^n)\in\mathcal{T}_{[VX_1X_2Y]}^n$.
\end{enumerate}
\end{enumerate}
Thus, the probability of error $\textsf{Pr}[E]$ can be upper bounded as
\begin{align}
&\textsf{Pr}[E_{0}]+\textsf{Pr}[E_1|E_{0}^c]
+\textsf{Pr}[E_{21}|E_1^c\cap E_0^c]
+\textsf{Pr}[E_{22}|E_1^c\cap E_0^c]+\textsf{Pr}[E_{23}|E_1^c\cap E_0^c]
+\textsf{Pr}[E_{24}|E_1^c\cap E_0^c]. \nonumber
\end{align}
We start with upper bounding $\textsf{Pr}[E_{0}]$. From the mutual covering lemma \cite{ElGamal:2011}, we have that 
$\textsf{Pr}[E_0]\le \epsilon$ if $r_1+r_2> I(X_1;X_2|V)$. Therefore, if the inequality in (\ref{Rconstraint01}) is satisfied, we have $\textsf{Pr}[E_0]\le \epsilon$.


Due to the Asymptotic Equipartition Property (AEP), for sufficient large $n$, we have
\begin{align}
\textsf{Pr}[E_1|E_{0}^c]\le\epsilon. \label{replyhard2}
\end{align}
Next, we proceed to upper bound $\textsf{Pr}[E_{21}|E_1^c\cap E_0^c]$. 
We have
\begin{align}
\textsf{Pr}\left[E_{21}|E_0^c\cap E_1^c \right]
&\leq \sum_{l_1'\ne l_1: x_1^n(i,l_1')}\textsf{Pr}\big[\left(v^n(i),x_1^n(i,l_1'),x_2^n(i,l_2),Y^n\right)
 \in\mathcal{T}_{[VX_1X_2Y]}^n   \big|E_1^c \cap E_0^c,  \big]  \nonumber\\
 &\le2^{-n(I(X_1;Y,X_2|V)-\epsilon)}2^{nR_1}.
\end{align}
which goes to zero because of (\ref{Rconstraint03}). Due to symmetry, the probability $\textsf{Pr}[E_{22}|E_1^c\cap E_0^c]$ goes to zero because of (\ref{Rconstraint02}).
The probability $\textsf{Pr}[E_{23}|E_1^c\cap E_2^c]$ can be bounded as follows
\begin{align}
&\textsf{Pr}[E_{23}|E_1^c\cap E_0^c] \nonumber\\
&=\sum_{l_1'\ne l_1,l_2'\ne l_2, (x_1^n(i,l_1'),x_2^n(i,l_2')) \in\mathcal{C}(i)}\textsf{Pr}
[\left(v^n(i),x_1^n(i,l_1'),x_2^n(i,l_2'),Y^n\right) \in\mathcal{T}^n_{[VX_1 X_2 Y]}|E_1^c\cap E_0^c]\nonumber\\
&\le 2^{-n(I(X_1,X_2;Y|V)-\epsilon)}2^{n(R_1-r_1+R_2-r_2)}\nonumber\\
&\le 2^{-n(I(X_1,X_2;Y|V)-\epsilon)}2^{n(R_1+R_2-I(X_1;X_2|V)-\delta)}.\label{appscc}
\end{align}
where the inequality in (\ref{appscc}) is due to (\ref{Rconstraint01}) and the quantities in (\ref{appscc})
goes to zero due to (\ref{Rconstraint05}).

The probability $\textsf{Pr}[E_{24}|E_1^c\cap E_2^c]$ can be bounded as follows
\begin{align}
&\textsf{Pr}[E_{24}|E_1^c\cap E_0^c]\nonumber\\
&=\sum_{i'\ne i,l_1',l_2' (x_1^n(i',l_1'),x_2^n(i',l_2'))
\in\mathcal{C}(i')}\textsf{Pr}[\left(v^n(i'),x_1^n(i',l_1'),x_2^n(i',l_2'),Y^n\right) \in\mathcal{T}^n_{[V X_1 X_2 Y]}|E_1^c\cap E_0^c]\nonumber\\
&\le 2^{-n(I(X_1,X_2;Y)-\epsilon)}2^{nR_0}2^{n(R_1-r_1+R_2-r_2)}\nonumber\\
&\le2^{-n(I(X_1,X_2;Y)-\epsilon)}2^{n(R_0+R_1+R_2-I(X_1;X_2|V)-\delta)}. \label{AT2}
\end{align}
which goes to zero due to (\ref{Rconstraint04}). Thus, as long as the rate tuple satisfies (\ref{Rconstraint01})-(\ref{Rconstraintd}), the probability of error can be made arbitrarily small, and rate $R$ according to (\ref{Rconstrainte}) is achievable. Using Fourier-Motzkin elimination and noting the fact that the capacity is defined as the \emph{supremum} of all achievable rates, we obtain the result stated in Theorem \ref{achiev}. 
%
%
%
%
%
\subsection{Proof of Theorem \ref{capacity}} \label{proof_capacity}
 We first show that if the Gaussian multiple access diamond channel satisfies (\ref{cond1}) and (\ref{cond2}), then $\bar{\rho}_3$
and $\bar{\rho}_5$ always exist.  Since the channel satisfies
(\ref{cond2}), we have $f_3(0)>f_4(0)$ and $f_5(0)>f_4(0)$. Since
the channel satisfies (\ref{cond1}), we have $f_5(1)<f_4(1)$.
Furthermore, we always have $f_3(1)<f_4(1)$. Since all functions
$f_3$, $f_4$ and $f_5$ are continuous, by the intermediate value
theorem, $\bar{\rho}_3$ and $\bar{\rho}_5$ always exist. Furthermore, since
$f_3(\rho)$, $f_4(\rho)$ and $f_5(\rho)$ are all monotonic, the root $\bar{\rho}_k \in [0,1]$, $k=3,5$ is unique.

\begin{figure}[t!]
\centering
\includegraphics[width=7in]{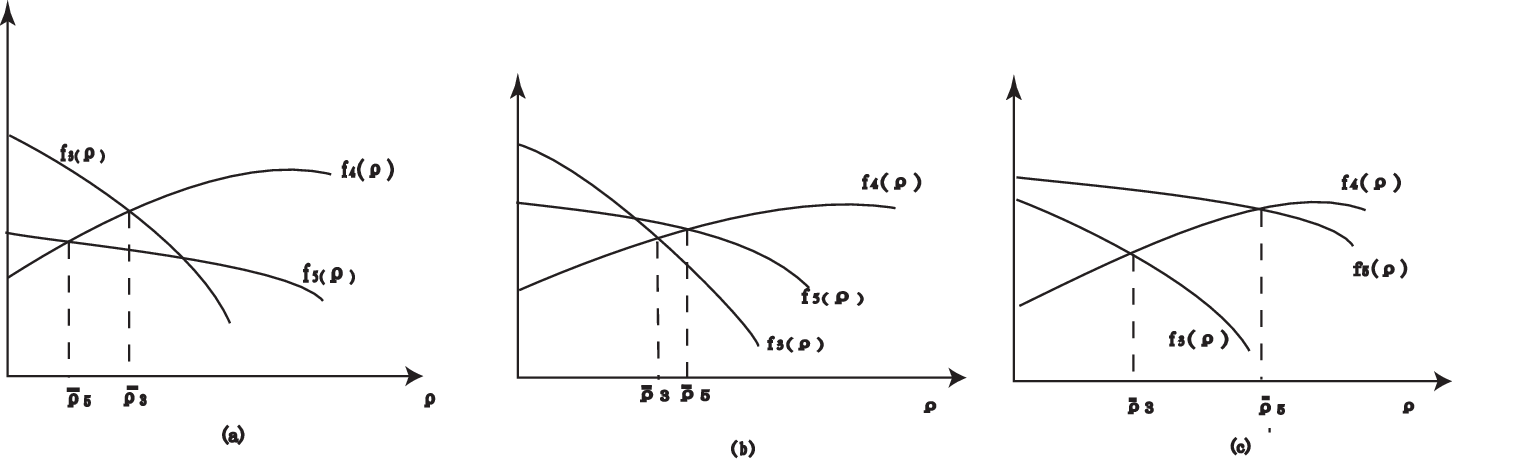}
\caption{The possible cases for $f_3$, $f_4$ and $f_5$.}
\label{CWW_cases}
\end{figure}

It is clear that $f_3(\rho)$ and $f_5(\rho)$ are strictly decreasing
in $\rho$, while $f_4(\rho)$ is strictly increasing in $\rho$, for
$\rho\in[0,1]$.
We also know that $f_3(1)<f_5(1)$. Noting that both $f_3(\rho)$ and $f_5(\rho)$ are functions of $\rho$ in terms of $1-\rho^2$ only, it can be shown that $f_3(\rho)=f_5(\rho)$ have at most one root in $\rho \in [0,1]$. 
%
%
%
%
Thus, if the
channel further satisfies $f_3(0)>f_5(0)$, then $f_3$, $f_4$ and
$f_5$ would look like either Figure \ref{CWW_cases} (a) or Figure
\ref{CWW_cases} (b). Otherwise, $f_3$, $f_4$ and $f_5$ would look
like Figure \ref{CWW_cases} (c).

We discuss the following cases:
\begin{enumerate}
\item In the case of Figure \ref{CWW_cases} (a), $f_3(0)>f_5(0)$ and $\bar{\rho}_5 \leq
\bar{\rho}_3$. In this scenario,
the lower bound is $f_4(\bar{\rho}_5)$. The cut-set bound is also $f_4(\bar{\rho}_5)$. Thus, in this case, we have $C_{\text{lower}}=C_{\text{upper}}=C_{\text{cut}}$. 

\item In the case of Figure \ref{CWW_cases} (b), $f_3(0)>f_5(0)$ and $\bar{\rho}_3 <
\bar{\rho}_5$. In this scenario,
the lower bound is $f_4(\bar{\rho}_3)$. The cut-set bound is $f_4(\bar{\rho}_5)$. As for the upper bound, we have the following sub-cases: 
\begin{enumerate}
\item if $\rho^* \leq
\bar{\rho}_3$, then $T_1=f_4(\rho^*)$ and $T_2=f_4(\bar{\rho}_5)$.
Since $T_1 < T_2$, the upper bound is $f_4(\bar{\rho}_5)$. Thus, in this case, we have $C_{\text{lower}}<C_{\text{upper}}=C_{\text{cut}}$. 
\item If $\bar{\rho}_5 \geq \rho^* > \bar{\rho}_3$, then
$T_1=f_4(\bar{\rho}_3)$, $T_2=f_4(\bar{\rho}_5)$ and $T_1 < T_2$,
the upper bound is $f_4(\bar{\rho}_5)$. Thus, in this case, we have $C_{\text{lower}}<C_{\text{upper}}=C_{\text{cut}}$. 
\item If $\rho^*> \bar{\rho}_5$, then $T_1=f_4(\bar{\rho}_3)$,
$T_2=f_5(\rho^*)$ and the upper bound is $\max (f_4(\bar{\rho}_3),
f_5(\rho^*))$. Thus, we have the following 2 sub-cases:
\begin{enumerate}
\item $f_5(\rho^*) \leq
f_4(\bar{\rho}_3)$. In this case, we have we have $C_{\text{lower}}=C_{\text{upper}}<C_{\text{cut}}$. 
\item $f_5(\rho^*) >
f_4(\bar{\rho}_3)$. In this case, we have $C_{\text{lower}}<C_{\text{upper}}<C_{\text{cut}}$. 
\end{enumerate}


\end{enumerate}
\item In the case of Figure \ref{CWW_cases} (c), $f_3(0) \leq f_5(0)$. Since we have $f_3(1) < f_5(1)$, we have that $\bar{\rho}_3 < \bar{\rho}_5$ in this case. The lower bound is $f_4(\bar{\rho}_3)$. The cut-set bound is $\min(f_3(0), f_4(\bar{\rho}_5))$. As for the upper bound, we have the following sub-cases:
\begin{enumerate}
\item if $\rho^*
\leq \bar{\rho}_3$, then $T_1=f_4(\rho^*)$ and $T_2=\min(f_3(0),
f_4(\bar{\rho}_5))$. Since $T_1< T_2$, the upper bound is
$\min(f_3(0), f_4(\bar{\rho}_5))$. Thus, in this case, we have $C_{\text{lower}}<C_{\text{upper}}=C_{\text{cut}}$. 
\item If $\bar{\rho}_3 < \rho^* \leq \bar{\rho}_5$,
then $T_1=f_4(\bar{\rho}_3)$, $T_2=\min(f_3(0), f_4(\bar{\rho}_5))$. Note that
$
f_4(\bar{\rho}_3)=f_3(\bar{\rho}_3)<f_3(0)
$.
Thus, $T_1 < T_2$, the upper bound is $\min(f_3(0),
f_4(\bar{\rho}_5))$. Thus, in this case, we have $C_{\text{lower}}<C_{\text{upper}}=C_{\text{cut}}$.
\item If $\rho^*> \bar{\rho}_5$, then $T_1=f_4(\bar{\rho}_3)$,
$T_2=\min(f_3(0),f_5(\rho^*))$ and the upper bound is $\max
(f_4(\bar{\rho}_3), \min(f_5(\rho^*), f_3(0)))$. we further have the following sub-cases:
\begin{enumerate}
\item 
$f_5(\rho^*) \leq f_4(\bar{\rho}_3)$, we have $C_{\text{lower}}=C_{\text{upper}}<C_{\text{cut}}$.  
\item $f_5(\rho^*) > f_4(\bar{\rho}_3)$, then the upper bound is $\min(f_5(\rho^*), f_3(0))$, we further have the following cases:
\begin{enumerate}
\item $f_3(0) \leq f_5(\rho^*)$. In this case, $C_{\text{upper}}=f_3(0)$. Since we are considering the case of $\rho^* > \bar{\rho}_5$, and due to the fact that $f_5(\cdot)$ is a decreasing function, we have $C_{\text{cut}}=f_3(0)$.  Thus, in this case, we have $C_{\text{lower}}<C_{\text{upper}}=C_{\text{cut}}$.
\item $f_3(0) > f_5(\rho^*)$. In this case, $C_{\text{upper}}=f_5(\rho^*)$, and we have $C_{\text{lower}}<C_{\text{upper}}<C_{\text{cut}}$.
\end{enumerate}
\end{enumerate}
\end{enumerate}
\end{enumerate}

Combining the result for these cases, and noting that we have $\bar{\rho}_5 \leq \bar{\rho}_3$ implies $f_3(0)>f_5(0)$, $f_5(\rho^*) \leq f_4(\bar{\rho}_3)$, together with $\bar{\rho}_3<\bar{\rho}_5$, implies $\rho^* > \bar{\rho}_5$, $f_3(0)>f_5(0)$ implies $f_3(0) >f_5(\rho^*)$, $f_5(\rho^*) \leq f_4(\bar{\rho}_3)$ implies $f_5(\rho^*) < f_3(0)$, we obtain the result of Theorem \ref{capacity}.

\section{Conclusions}
We have studied the Gaussian multiple access diamond channel. Noting
the similarity between this problem and the Gaussian multiple
description problem, we first provide an upper bound on the
capacity. We then obtain an achievable rate by correlated code with the superposition structure. Finally, we provide conditions that our proposed upper and lower bounds meet. Thus,
for a
 Gaussian multiple access diamond channel
 that satisfies these conditions, we have found its capacity.



%
\bibliographystyle{unsrt}
\bibliography{/Users/wkang/Documents/Dropbox/writing/refphd}

\begin{thebibliography}{10}

\bibitem{Schein:2001}
B.~E. Schein.
\newblock {\em Distributed Coordination in Network Information Theory}.
\newblock PhD thesis, Massachusetts Institute of Technology, 2001.

\bibitem{Cover:1991}
T.~M. Cover and J.~A. Thomas.
\newblock {\em Elements of Information Theory}.
\newblock John Wiley and Sons, 1991.

\bibitem{Kang:2011b}
W.~Kang and S.~Ulukus.
\newblock Capacity of a class of diamond channels.
\newblock {\em IEEE Trans. Inform. Theory}, 57:4955--4960, Aug. 2011.

\bibitem{Tandon:2010}
R.~Tandon and S.~Ulukus.
\newblock Diamond channel with partially separated relays.
\newblock In {\em Proc. IEEE International Symp. on Information Theory (ISIT)},
  pages 644--648, Austin, TX, June 2010.

\bibitem{Avestimehr:2011}
A.~S. Avestimehr, S.~N. Diggavi, and D.~N.~C. Tse.
\newblock Wireless network information flow: A deterministic approach.
\newblock {\em IEEE Trans. Inform. Theory}, 57(4):1872--1905, Apr. 2011.

\bibitem{Niesen:2013}
U.~Niesen and S.~N. Diggavi.
\newblock The approximate capacity of the {G}aussian {N}-relay diamond network.
\newblock {\em IEEE Trans. Inform. Theory}, 59(2):845--859, Feb. 2013.

\bibitem{Ahlswede:1983}
R.~Ahlswede and T.~S. Han.
\newblock On source coding with side information via a multiple-access channel
  and related problems in multi-user information theory.
\newblock {\em IEEE Trans. Inform. Theory}, 29(3):396--412, 1983.

\bibitem{Traskov:2007}
D.~Traskov and G.~Kramer.
\newblock Reliable communication in networks with multi-access interference.
\newblock In {\em Proc. Conf. IEEE Information Theory Workshop (ITW)}, Lake
  Tahoe, CA, Sep. 2007.

\bibitem{Cover:1980}
T.~M. Cover, A.~El~Gamal, and M.~Salehi.
\newblock Multiple access channel with arbitrarily correlated sources.
\newblock {\em IEEE Trans. Inform. Theory}, 26:648--657, Nov. 1980.

\bibitem{Willems:1983}
F.~M.~J. Willems.
\newblock The discrete memoryless multiple access channel with partially
  cooperating encoders.
\newblock {\em IEEE Trans. Inform. Theory}, 29(3):441--445, May 1983.

\bibitem{Bross:2008}
S.~I. Bross, A.~Lapidoth, and M.~A. Wigger.
\newblock The {G}aussian {MAC} with conferencing encoders.
\newblock In {\em Information Theory, 2008. ISIT 2008. IEEE International
  Symposium on}, pages 2702--2706, 2008.

\bibitem{Marton:1979}
K.~Marton.
\newblock A coding theorem for the discrete memoryless broadcast channel.
\newblock {\em IEEE Trans. Inform. Theory}, 25:306--311, May 1979.

\bibitem{Ozarow:1980}
L.~Ozarow.
\newblock On a source-coding problem with two channels and three receivers.
\newblock {\em Bell Syst. Tech. J.}, 59:1909--1921, December 1980.

\bibitem{Bergmans:1974}
P.~Bergmans.
\newblock A simple converse for broadcast channels with additive white
  {G}aussian noise.
\newblock {\em IEEE Trans. on Information Theory}, 20:279 --280, March 1974.

\bibitem{Bidokhti:2014}
S.~S. Bidokhti and G.~Kramer.
\newblock Capacity bounds for a class of diamond networks.
\newblock {\em arXiv preprint arXiv:1401.6135}, 2014.

\bibitem{Csiszar:1981}
I.~Csiszar and J.~Korner.
\newblock {\em Information Theory: Coding Theorems for Discrete Memoryless
  Systems}.
\newblock Academic Press, 1981.

\bibitem{ElGamal:2011}
A.~El Gamal and Y.~H. Kim.
\newblock {\em Network information theory}.
\newblock Cambridge University Press, 2011.

\end{thebibliography}
\end{document}